\documentclass[aps,physrev,twocolumn,groupedaddress,amsmath,amssymb,10pt]{revtex4-2}

\usepackage{graphicx}
\usepackage{hyperref}
\usepackage{physics}

\newcommand{\Op}[1]{\hat{\mathrm{#1}}}

\begin{document}


\title{Chiral cavity-induced quantum phase transitions in a quantum ring.}


\author{Loic Remolif}
\email[]{loic.remolif@queensu.ca}
\author{\href{https://orcid.org/0000-0003-4992-6122}{Ivan Iorsh}}
\email[]{ivan.iorsh@queensu.ca}
\affiliation{Department of Physics, Engineering Physics and Astronomy, Queen's University, Kingston, ON, Canada}



\date{\today}

\begin{abstract}
We consider a quantum ring placed in a gyrotropic cavity characterized by the energy splitting between the left and right circularly polarized modes. We show that despite the absence of constant magnetic field penetrating through the ring, in the regime of the ultrastrong light matter coupling, the total current in the ground state changes discontinuously with the light matter coupling in the direct analogy with the Aharonov-Bohm ring. We consider the driven-dissipative of the system and show that the discontinuous change of the total angular momentum can be directly probed via the spectral and statical properties of the radiation emitted by the system under weak coherent drive.
\end{abstract}


\maketitle

\section{Introduction}
The interaction of light with matter with the aim of manipulating the properties of the latter has been extensively studied and remains an active and rapidly evolving topic both in physics~\cite{Mivehvar_2021,Rokaj_2022}, especially condensed matter and quantum optics, and in chemistry~\cite{Mandal_2023,Hsu_2025}. Light-matter interactions are of particular importance in the strong regime coupling, where the rate of energy exchange between the two subsystems is greater than dissipative effects~\cite{Barnes_2015,Bahsoun_2018,Frisk_K_2019,Lechner_2023}, and cavity quantum electrodynamics (cavity QED or cQED) offers a convenient and efficient platform to obtain this environment in a nonrelativistic regime, by confining light in small volumes.

In the extreme regime of light matter coupling, termed ultrastrong coupling regime~\cite{Frisk_K_2019}, the characteristic energy of light matter interaction becomes non-negligible as compared to the cavity photon energy. In this regime, vacuum fluctuations may substantially modify the properties of the ground state of the material system even in the absence of external illumination. In recent years, cavity QED in the ultrastrong coupling regime has been predicted to create, enhance, or simply mediate material properties, such as magnetism~\cite{Roman_roche_2021}, many-body localization~\cite{Ge_2025}, superconducting phase~\cite{Schlawin_2019,Young_2024,I_te_2024,Kozin_2025}, and other phenomena~\cite{GarciaVidal_2021}. The role of cavity modified vacuum fluctuations of electromagnetic field was first revealed in the Casimir effect~\cite{Casimir_1948,Lamoreaux_1997}, where the discretization of the cavity photon eigenfrequencies leads to the emergence of the attractive force between the cavity mirrors.

While the first works on the ultrastrong coupling regime in cavity QED considered the single mode cavity case, it soon became apparent that engineering the cavity mode structure opens a new degree of freedom in tailoring the ground state properties of the cavity embedded materials. A particularly interesting class of cavities are gyrotropic cavities~\cite{Hubener_2021}, where breaking of the time reversal symmetry induces energy splitting between the cavity eigenmodes of two circular polarizations. It has been previously shown that circularly polarized vacuum fluctuations in such cavities can play the role of the magnetic field for the cavity embedded material even when there is neither magnetic field directly interacting with matter nor magnetic flux penetrating through the sample. Specifically, a number of cavity induced effects were predicted including the classical Hall effect~\cite{Tokatly_2021,Sedov_2022}, quantum integer~\cite{Appugliese_2022} and fractional~\cite{Bacciconi_2025} Hall effects, photon condensation~\cite{Zeno_2023}, generation of entangled states~\cite{Passetti_2023}, topological phase transitions~\cite{Zeno_2024} and peculiar optomechanical phenomena~\cite{Ilin_2024}.

\begin{figure}
\includegraphics[width=.73\linewidth]{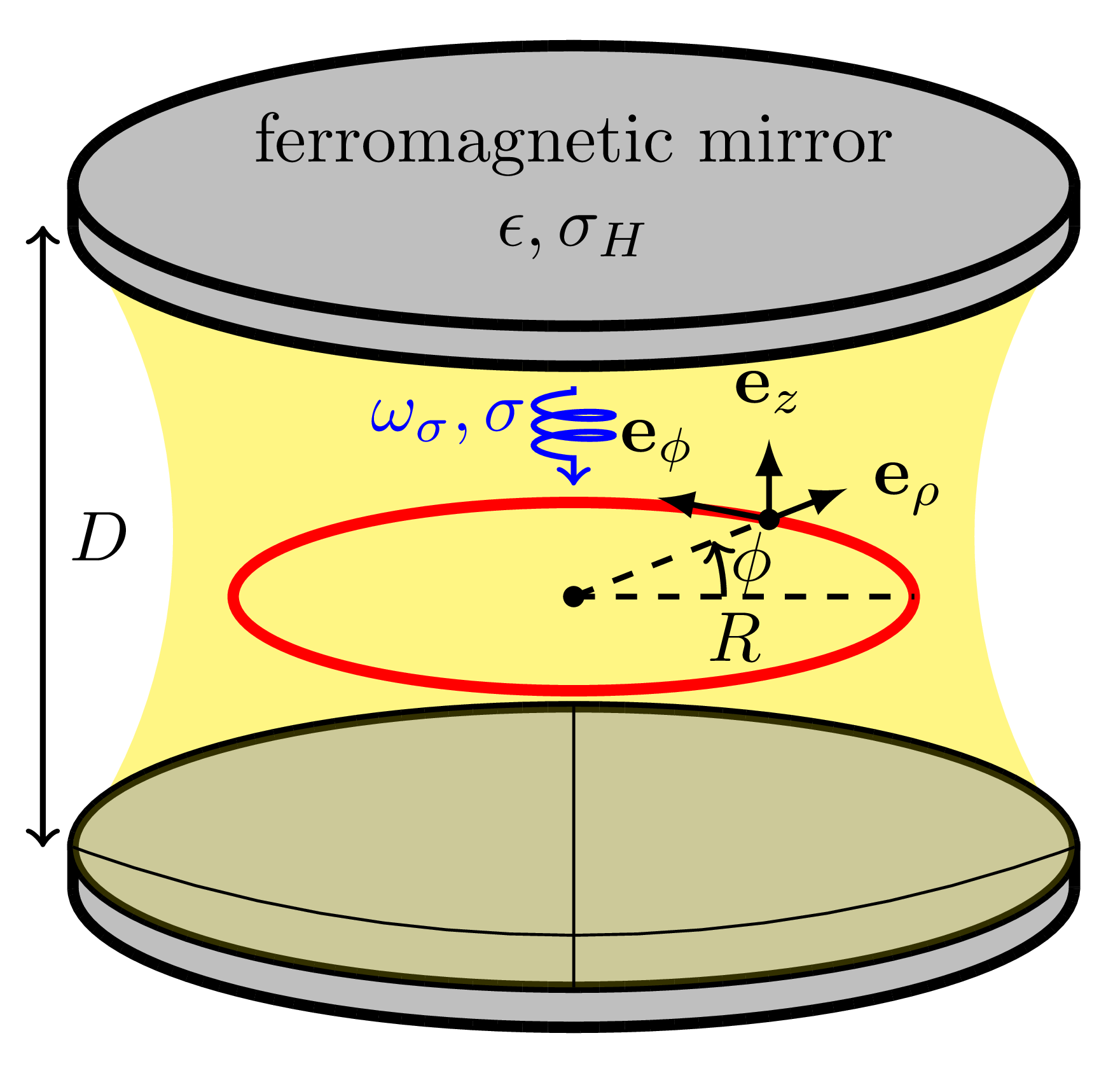}
\caption{\label{fig:cavity}One-dimensional ring of radius $R$ (red) embedded in a Fabry-Perot chiral cavity of length $D$ formed by ferromagnetic mirrors in the presence of photons (blue) with energy $\hbar\omega_{\sigma}$ and circular polarization $\sigma=\pm$ propagating in the $z$-direction.}
\end{figure}

In this paper, we consider the system schematically shown in Fig.~\ref{fig:cavity}: a quantum ring is placed inside a gyrotropic cavity. We first show that there exists a threshold light matter coupling strength, above which the ground state of the ring supports the persistent current. We show that the dependence of the current on the coupling strength is step-like similar to the Aharonov-Bohm effect, despite the fact that there is neither constant magnetic flux penetrating the ring nor circularly polarized illumination. We then consider the system in equilibrium with a thermal bath and build the phase diagram of the ground state current. Finally, we consider the stationary states of the continuously driven cavity and show the entanglement between of the states corresponding to different angular momenta.

The emergence of persistent currents in quantum rings embedded in gyrotropic cavity was considered in~\cite{Kibis_2013}. In this work, the multimode nature of the cavity and the weak coupling regime precluded the emergence of the quantum phase transition of the ground state and thus no step like dependence of the current on the light matter coupling was considered, which is a primary focus of the current work.

\section{Model and method}
In this section, we present the Hamiltonian of the system schematically shown in Fig.~\ref{fig:cavity}. We consider the minimal coupling Hamiltonian yielding (hereafter we set $\hbar=1$):
\begin{equation}
    \Op{H} = \frac{1}{2m}\left(\Op{\mathbf{p}}-\frac{e}{c}\Op{\mathbf{A}}\right)^2 +V\left(\mathbf{r}\right) + \sum_{\sigma=\pm}\omega_{\sigma}\left(\Op{a}_{\sigma}^{\dagger}\Op{a}_{\sigma} + \frac{1}{2}\right),
\end{equation}where $\mathbf{p,r}$ are the momentum and position of the electron, $V(\mathbf{r})$ is the confining potential, $\sigma=\pm$ is the index of the circular polarization, and $\omega_{+}\neq \omega_{-}$ due to the gyrotropy in the cavity. We only consider two lowest cavity modes and assume long wavelength limit i.e. that the field profiles are spatially uniform at the scale of ring radius. The splitting of two circularly polarized modes requires the breaking of time-reversal symmetry. Such a gyrotropic cavity can be realized for instance via a Fabry-Perot cavity with mirrors made of ferromagnetic material (see App.~\ref{app:gyrocac} for details) or by more complex geometries such as gyrotropic metamaterials~\cite{Zhang_2009,Wang_2009,Wang_2016,Wang_2022,Begum_2023,Xu_2024}. 


In the case of a sufficiently narrow ring, we can consider only the lowest electron eigenstate in the radial transverse direction and project the Hamiltonian to the one-dimensional one (see App.~\ref{app:Heff} for details):
\begin{equation}
    \Op{H} = \frac{1}{2mR^2}(-i\partial_{\phi}-\frac{e}{c}R\Op{A}_{\phi})^2 +\sum_{\sigma=\pm}\omega_{\sigma}\left(\Op{a}_{\sigma}^{\dagger}\Op{a}_{\sigma} + \frac{1}{2}\right),
\end{equation}where $R$ is the radius of the ring. The azimuthal component of the vector potential $\Op{A}_{\phi}$ reads
\begin{equation}
    \frac{e}{c}R\Op{A}_{\phi} = \sum_{\sigma=\pm}g_{\sigma}\sigma i (\Op{a}_{\sigma}e^{i\sigma\phi} - \Op{a}_{\sigma}^{\dagger}e^{-i\sigma\phi}),
\end{equation}
where $g_{\sigma}=\sqrt{e^2 R^2\pi/(V \omega_{\sigma})}$ is the dimensionless light matter coupling, and $V$ is the effective cavity volume. We normalize the energy to $1/(2mR^2)$ and introduce dimensionless cavity frequencies $\tilde{\omega}_{\sigma}= \omega_{\sigma}(2mR^2)$. We further introduce the dimensionless parameter $\xi = V/R^3$. For the conventional cavity $\xi \gg 1$, however, for deeply subwavelength cavities $\xi$ can become of the order of unity. The dimensionless Hamiltonian then yields
\begin{equation}
    \Op{H} = \left(-i\partial_{\phi}-g_{0}\sum_{\sigma=\pm}\tfrac{i\sigma (\Op{a}_{\sigma}e^{i\sigma\phi}-\mathrm{h.c.})}{\sqrt{\tilde{\omega}_{\sigma}}}\right)^{2} + \sum_{\sigma=\pm}\tilde{\omega}_{\sigma}\left(\Op{n}_{\sigma}+\tfrac{1}{2}\right),
\end{equation}where $\Op{n}_{\sigma}=\Op{a}_{\sigma}^{\dagger}\Op{a}_{\sigma}$ and $g_{0}=\sqrt{2\pi R/(a_B\xi)}$, where $a_B=\hbar^2/(me^2)$ is the Bohr radius.

In order to remove the $\phi$-dependence from the Hamiltonian, we make the unitary transformation $\Op{H}\to\Op{U}\Op{H}\Op{U}^{\dagger}$, where 
\begin{equation}
    \Op{U} = \exp\left[i\sum_{\sigma=\pm} \sigma \Op{n}_{\sigma}\phi\right].
\end{equation}We further apply the displacement operator:
\begin{equation}
    \Op{D} = \exp\left[\sum_{\sigma=\pm}\left(ig_{\sigma}\Op{a}_{\sigma}+\mathrm{h.c.}\right)\right].
\end{equation}
The transformed Hamiltonian $\Op{H}^{\prime}$ is then written as:
\begin{align}
    \Op{H}^{\prime} =& \left(-i\partial_{\phi}+\Phi+\Op{n}_{-}-\Op{n}_{+}\right)^2+\nonumber\\
    &\sum_{\sigma=\pm}\omega_{\sigma}\left[(\Op{a}_{\sigma}^{\dagger}-ig_{\sigma})(\Op{a}_{\sigma}+ig_{\sigma})+\frac{1}{2}\right]\label{eq:Hprime},
\end{align}
where $\Phi=g_+^2-g_-^2$. We can see that the transformed commutes with $\partial_{\phi}$, which reflects the conservation of total angular momentum of the ring+cavity system.
\begin{figure}[!h]
\includegraphics[width=\linewidth]{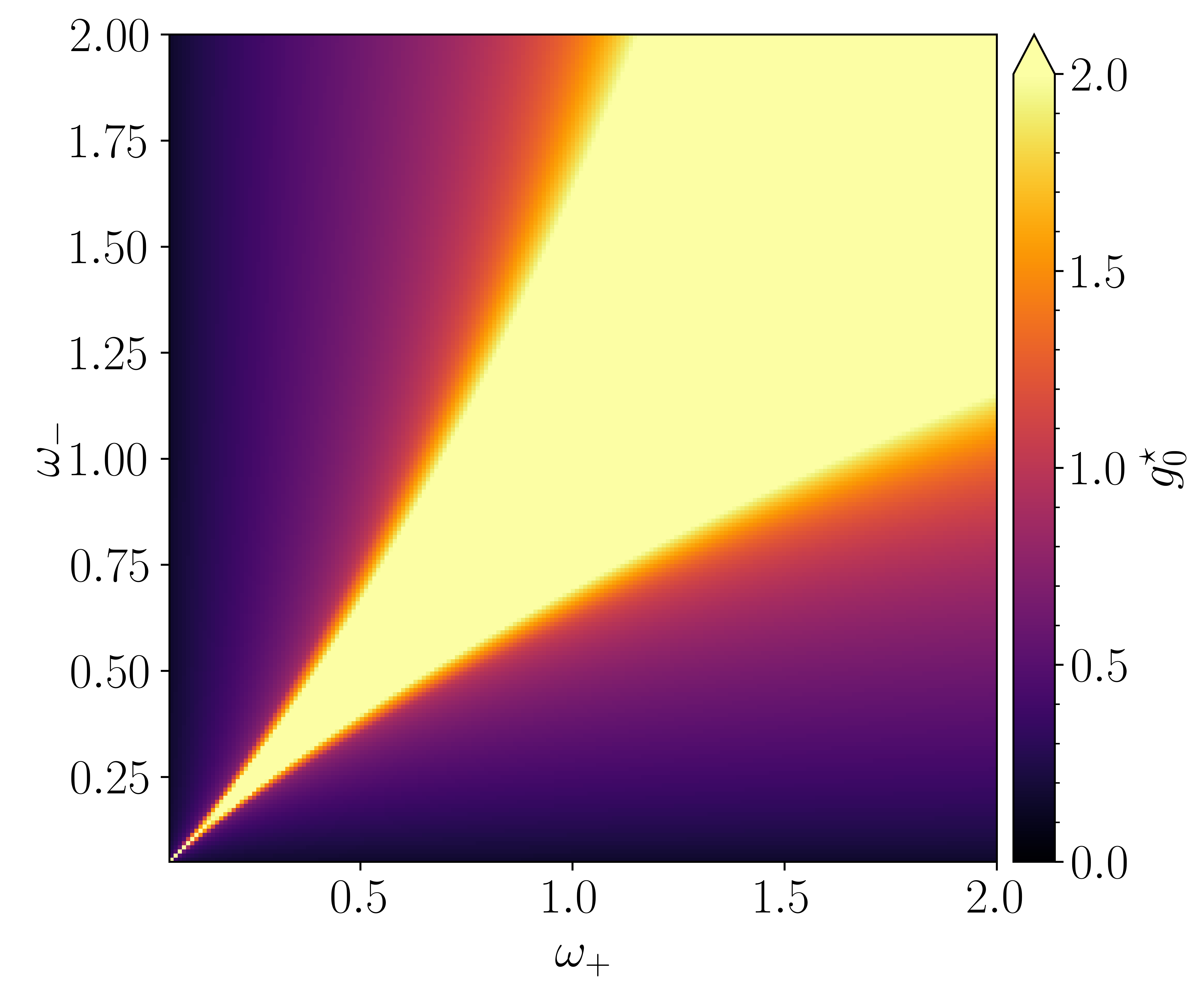}
\caption{\label{fig:critical_g2}Critical light-matter coupling $g_{0}^{\star}$ at which a transition $j=0\rightarrow j=\pm1$ occurs as a function of the frequency $\omega_{\pm}$. Low values of $g_{0}^{\star}$ are obtained for large detuning, while keeping one frequency small.}
\end{figure}

The full Hamiltonian can then be separated into independent blocks corresponding to different total angular momentum $j$ of the system $-i\partial_{\phi}\rightarrow j=0,\pm 1,\pm 2\ldots$. For zero and small $g_{0}$ the lowest energy state always corresponds to $j=0$, however, for some critical value of $g_{0}$, $g_{0}^{\star}$ the lowest energy state with $j=\pm 1$ intersects the state with $j=0$ and for larger $g_{0}$ the system is characterized with non zero total angular momentum.
\section{Results}
In Fig.~\ref{fig:critical_g2} we show the dependence of the critical light matter coupling $g_{0}^{\star}$ on the cavity mode frequencies $\omega_{\pm}$. We can see that the lowest values are achieved when one of the frequencies $\tilde{\omega}_\sigma \gg 1$ and the opposite one $\omega_{\bar{\sigma}} \ll 1$. As we show in the App.~\ref{app:sml} in the limiting case, when one of the frequencies, e.g. $\omega_{-}\gg 1$, we can project to the vacuum state of this mode yielding the Hamiltonian for the total angular momentum $j$ (neglecting the vacuum photon energy):
\begin{align}
\Op{H}^{\prime}_j = (j+g_+^2-\Op{n}_{+})^2+\omega_+(\Op{a}_{+}^{\dagger}-ig_{+})(\Op{a}_{+}+ig_{+}) \label{eq:1d}
\end{align}
We note that the Hamiltonian written in this form resembles that of the resonantly drive Kerr oscillator. Indeed, denoting $\tilde{j}=j+g_+^2$, we rewrite Eq.~\eqref{eq:1d}:
\begin{align}
\Op{H}^{\prime}_j=\tilde{j}^2 +\Op{n}_{+}(\Op{n}_{+}-(2\tilde{j}-\omega_+))+\omega_+ g_+i (\Op{a}_+^{\dagger}-\Op{a}_+)+g_+^2. \label{eq:H_reduced}
\end{align}
We note that in the regime where $\omega_+g_+ \ll 1$ and at the same time $2\tilde{j}-\omega \approx$ in finding the ground state, we can limit our consideration only to the first two Fock states $n_{+}=0,1$. The Hamiltonian projected to the first two Fock states can be written as
\begin{align}
\Op{H}^{\prime}_j \approx \tilde{j}^2 + \frac{1-2\tilde{j}+\tilde{\omega}_+}{2}(1+\Op{\sigma}_z)+\tilde{g}_+\tilde{\omega}_+\Op{\sigma}_y,
\end{align}
where $\Op{\sigma}_z,\Op{\sigma}_y$ are the Pauli matrices in the basis of the first two Fock states. For each $\tilde{j}$ there are two eigenstates $|\psi_{j,L}\rangle$, $\psi_{j,U}\rangle$ corresponding to lower and higher energy. For $\omega_{+} \ll 1$ as $g_+$ approaches $1/\sqrt{2}$ transition occurs between the energies $E_{0,L}$ and $E_{-1,L}$ which results in the emergence of the non-vanishing total angular momenta in the ground state. In the vicinity of the transition, three states form an almost degenerate manifold, $|\psi_{0,L}\rangle, |\psi_{0,U}\rangle$ and $|\psi_{-1,L}\rangle$. 
\begin{figure}
\includegraphics[width=\linewidth]{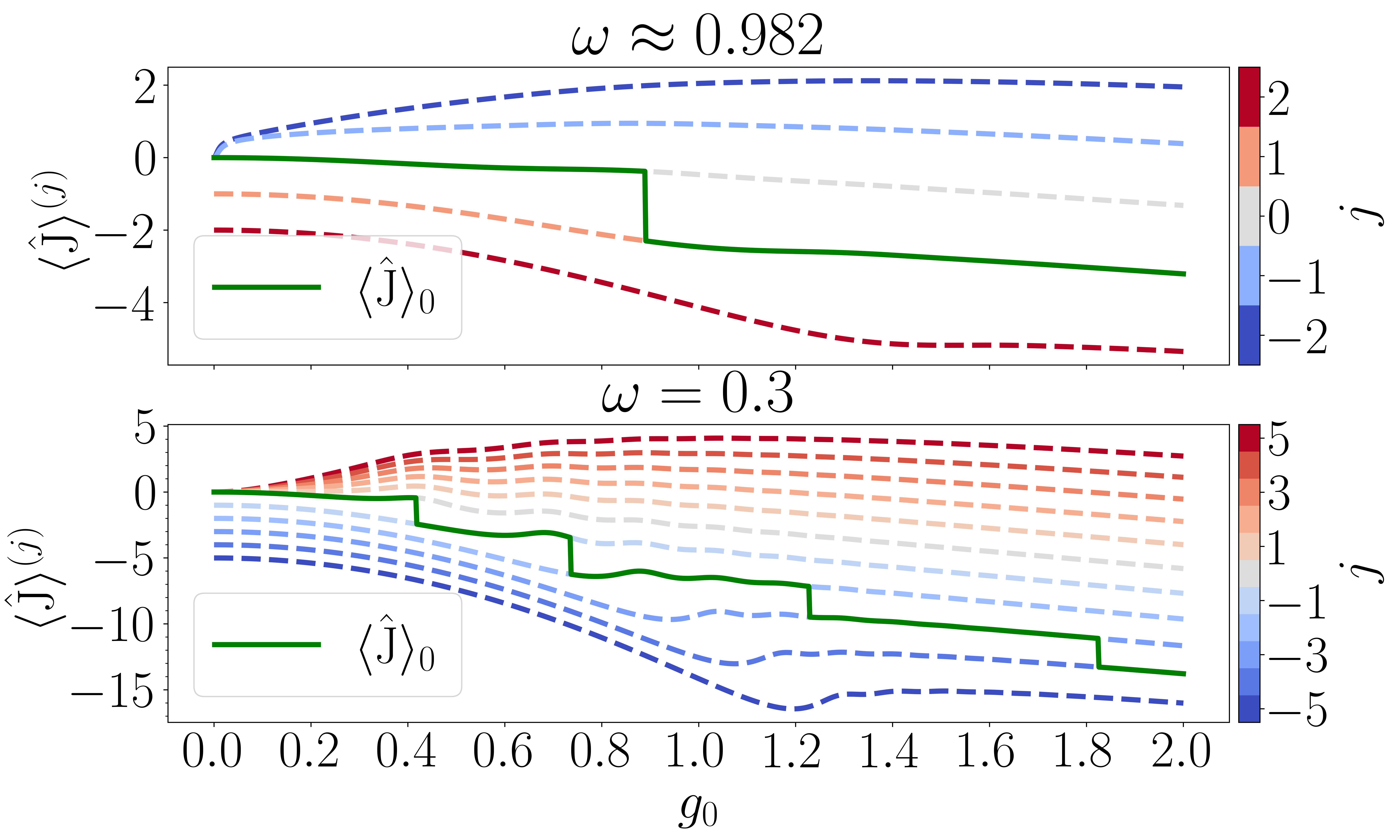}
\caption{\label{fig:current1}Ground state current $\ev{\operatorname{J}}^{(j)}$ as a function of the light-matter coupling $g_{0}$, for $\omega=5\pi/16\approx0.982$ (top) and $\omega=0.3$ (bot.) with $\sigma=+1$, for different total angular momenta $-5\le j\le 5$.}
\end{figure}
For even larger $g_{0}$, the next transition may occur when the ground state is characterized by the angular momentum $j=-2$. We can compute the value of the total dimensionless  current in the ground state which is given by $J= j_{GS}+g_+^2 -\langle \Psi_0 | n_+ |\Psi_0\rangle$, where $|\Psi_0\rangle$ is the ground state. In Fig.~\ref{fig:current1} we plot the dependence of the ground state current on the light-matter coupling $g_{0}$. We can see that as $g_{0}$ the system experiences several step-like changes in the ground state current similar to the Aharonov-Bohm effect. At the same time, the step heights are not universal since the cavity fluctuations are defined by the cavity geometry.

\begin{figure}[!h]
\includegraphics[width=\linewidth]{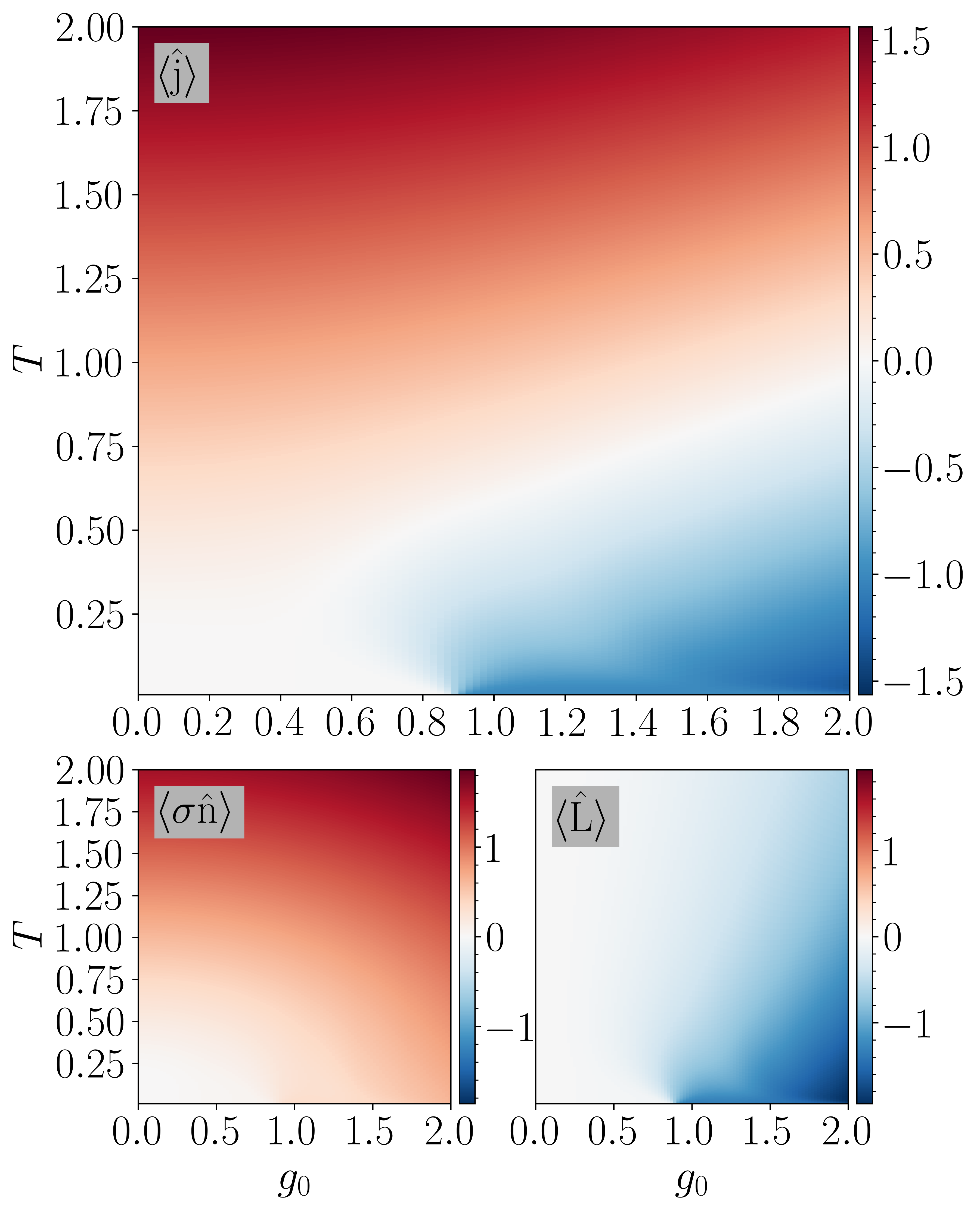}
\caption{\label{fig:QPT1}Quantum phase transition in the expectation value of the angular momentum as a function of the light-matter coupling $g_{0}$ and dimensionless temperature $T$ for $\omega=5\pi/16$, $\sigma=+1$. Top: total angular momentum $\langle\Op{j}\rangle$. Bottom left: photonic angular momentum $\langle\Op{J}_{p}\rangle$. Bottom right: electronic angular momentum $\langle\Op{L}\rangle$. The values would be inverted for $\sigma=-1$.}
\end{figure}

Furthermore, we can evaluate the ground state current in the case of finite temperature, when the system is in contact with a thermal bath. For finite temperature, the average current is given by
\begin{align}
J = \sum_{n} p_n \langle \Psi_n | \Op{j}+g_+^2-\Op{n}_{+})|\Psi_n\rangle,
\end{align}
where sum is taken over the eigenstates of the system, and $p_n$ are the thermal probabilities $p_n\sim e^{-E_n/(kT)}$. In Fig.~\ref{fig:QPT1} we plot the phase diagram of the current $J$ vs $g_{0}$ and temperature $T$.
The total angular momentum can be separated in the electronic part coming from the operator $\Op{j}$ and optical part coming from $\sum_{\sigma=\pm}\sigma \Op{n}_{\sigma}$. These two parts are plotted separately in Fig.~\ref{fig:QPT1}. As can be seen the optical angular momentum increases gradually as the temperature increases and a larger number of excited states become occupied. At the same time, the electronic angular momentum is almost identically zero for $g_{0}$ less than critical. For $g_{0}>g_{0}^{\star}$ the electronic angular momentum increases discontinuously at zero temperature and more gradually as the temperature increases. At finite temperature, the total angular momentum changes sign as the light-matter coupling passes through the critical light-matter coupling.

\begin{figure}
\includegraphics[width=\linewidth]{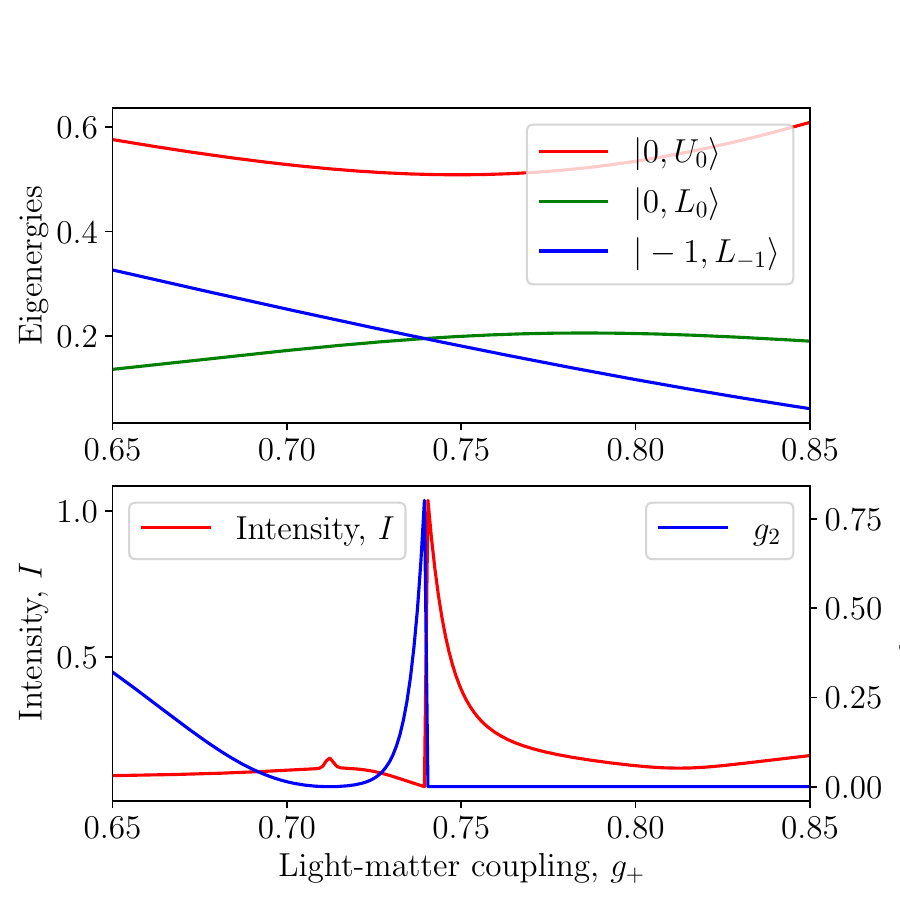}
\caption{\label{fig:autocorr} (Upper panel): dependence of the three lowest eigenergies of the closed system on the coupling parameter $g_+$. The value of $\tilde{\omega}_+=0.2$. (Lower panel) dependence of the output intensity and second order temporal correlation function $g_2(0)$ for the open system periodically driven with frequency $\omega_p=0.2$ and corresponding field amplitude $\Omega_p=0.6$. The decay rate is $\gamma=0.1$.}
\end{figure}

Finally, we consider the driven dissipative dynamics in the cavity subject to the coherent pump. Specifically, we supplement the Hamiltonian with the photon reservoir Hamiltonian $\Op{H}_{R}$, linear system-reservoir coupling $\Op{H}_{S-R}$ and the coherent pump $\Op{H}_{P}$, yielding:
\begin{align}
&\Op{H}_{R}= \sum_{k}\omega_k \Op{b}_{k}^{\dagger}\Op{b}_{k},\\
&\Op{H}_{S-R} = \sqrt{\gamma} \sum_k i(\Op{a}_{+}^{\dagger}b_k-\mathrm{h.c.}), \\
&\Op{H}_{P} = i\Omega_p   (\Op{a}^{\dagger}_{+}-\Op{a}_{+})\sin \omega_p t.
\end{align}
Here, $\Op{b}_k$ are the annihilation operators for the reservoir photons, $\Omega_p$ is proportional to the driving field amplitude, and $\omega_p$ is the pump frequency. In the limit of $g_+\approx 1/2$ and $\tilde{\omega}_+\ll 1$ we can limit the consideration to the first four lowest  levels corresponding to the eigenstates of Hamiltonian  $H_j$ in Eq.~\eqref{eq:H_reduced} for $j=0$ and $j=-1$. We label the four eigenstates as $|0,L_0\rangle$, $|0,U_0\rangle$, $|-1,L_{-1}\rangle$, $|-1,U_{-1}\rangle$ where $L,U$ correspond to the lower and upper eigenstate. Moreover, in the vicinity of the phase transition, the state $|-1,U_{-1}\rangle$ is detuned from the remaining three states. We therefore limit the computational basis to three states $|0,L_0\rangle$, $|0,U_0\rangle$, $|-1,L_{-1}\rangle$. For $g_{0}<g_{0}^{\star}$ the ground state is $|0,L_0\rangle$ state and for $g_{0}>g_{0}^{\star}$  - $|-1,L_{-1}\rangle$. We note that unitary transformations $\Op{U}, \Op{D}$ transform annihilation operators as:
\begin{align}
\Op{\mathcal{A}}=(\Op{D}\Op{U})\Op{a}_+((\Op{D}\Op{U}))=(\Op{a}_++ig_+)e^{-i\phi}
\end{align}
Therefore, pump and decay to reservoir only couple the states with different values of $j$. If we limit ourselves to the basis of three lowest lying states, we can adopt the dressed density matrix equation approach in secular approximation~\cite{Settineri_2018}, which yields:
\begin{align}
&\dot{\rho}= -i[\hat{\tilde{\mathrm{H}}}+\hat{\tilde{\mathrm{H}}}_P,\rho] +\nonumber \\
&\gamma\sum_{\omega,\omega^{\prime}>0}\left[(\omega+\omega^{\prime})\Op{\mathcal{E}}_{\omega}^{+}\rho \Op{\mathcal{E}}_{\omega'}^{-} - \omega\Op{\mathcal{E}}_{\omega^{\prime}}^{-} \Op{\mathcal{E}}_{\omega}^{+}\rho - \omega^{\prime}\rho \Op{\mathcal{E}}_{\omega'}^{-} \Op{\mathcal{E}}_{\omega}^{+}\right],
\end{align}
which is written in the zero temperature approximation and for the case of the Ohmic bath. The frequencies $\omega,\omega'$ span the differences between the three eigenenergies, $\omega = \epsilon_{\beta}-\epsilon_{\alpha}$, $\alpha,\beta \in \{|0,L_0\rangle, |0,U_0\rangle, |-1,L_{-1}\rangle \}$. The operators $\Op{\mathcal{E}}^+_{\omega}$ yield
\begin{align}
\Op{\mathcal{E}}^+_{\omega}=\langle \alpha | i\hat{\mathcal{A}}-i\hat{\mathcal{A}}^{\dagger} |\beta\rangle  |\beta\rangle \langle\alpha |,\quad (\omega = \epsilon_{\beta}-\epsilon_{\alpha}, \omega >0), \label{eq:Eplus}
\end{align}
and $\Op{\mathcal{E}}^-_{\omega}=(\Op{\mathcal{E}}^+_{\omega})^{\dagger}$. The output radiation intensity is given by $ I=\mathrm{Tr}[\Op{\mathcal{E}}^-\Op{\mathcal{E}}^+\rho(t)]$, where $\Op{\mathcal{E}}^-=\sum_{\omega} \Op{\mathcal{E}}^-_{\omega}$. The second order autocorrelation function $g_2=\mathrm{Tr}[\Op{\mathcal{E}}^-\Op{\mathcal{E}}^-\Op{\mathcal{E}}^+\Op{\mathcal{E}}^+\rho(t)]/I^2$. 
The dependence of emission intensity as well as of $g_2$ autocorrelation function is shown in Fig.~\ref{fig:autocorr}. We can see that the $g_2$ plot exhibits discontinuity at the level crossing. The origin of the discontinuity is evident if we consider the structure of the $\Op{\mathcal{E}}^+$ operator. For $g_{0}<g_{0}^{\star}$ it can be written as:
\begin{align}
\Op{\mathcal{E}}^+\|_{g_{0}<g_{0}^{\star}} = C_1|0,U_0\rangle \langle -1,L_{-1}| +C_2|-1,L_{-1}\rangle \langle 0, L_{0}|,
\end{align}
where $C_1,C_2$ are complex valued constants. Therefore, the operator $(\Op{\mathcal{E}}^+)^2$ will have the only non-vanishing matrix element $|0,U_0\rangle \langle 0, L_0|$. At the same time, for $g_{0}>g_{0}^{\star}$ the operator $\Op{\mathcal{E}}^+$ reads:
\begin{align}
\Op{\mathcal{E}}^+\|_{g_{0}>g_{0}^{\star}} = C_1|0,U_0\rangle \langle -1,L_{-1}| +C_2^*|0,L_{0}\rangle \langle -1, L_{-1}|,
\end{align}
and therefore its square will vanish leading to zero autocorrelation. We therefore suggest that quantum phase transition in the cavity-QED systems in ultrastrong coupling regime could be probed effectively by measuring statistical properties of light scattered of such systems.

\section{Discussion}
It is instructive to evaluate the parameters of the cavity required to evaluate the considered effect. We considered the case, when the characteristic kinetic energy of the electron in the ring is larger than the photon resonant frequency. For the semiconducting quantum ring made of GaAs with radius 20 nm, the value of $\hbar^2/(2mR^2)$ is approximately 1.3 meV, which corresponds to the terahertz frequency range and wavelength $1$~mm. We also need a considerable splitting of frequencies of two circularly polarized modes which has been achieved in THz cavities using resonant coupling to the inter-Landau level transition in two-dimensional materials~\cite{PhysRevB.109.L161302}. For a relative frequency splitting of 0.1 one would require the ratio of the photon wavelength and cavity volume $\lambda^3/V$ of the order  of $10^9$. The close ratios of $10^8$ have been recently demonstrated in ultrasubwavelength THz cavities~\cite{aupiais2023ultrasmall}.

Since gyrotropic cavity requires breaking of time reversal symmetry and thus the presence of magnetic field, the electron in the ring would be affected by this field, and it is required to discriminate the effect of constant field from the effect of vacuum fluctuations. It can be achieved by a tunable cavity in which the cavity volume can be adjusted changing only the strength of light matter coupling but not affecting the strength of the constant magnetic field.

\section{Conclusion}
We have shown that vacuum fluctuations in a gyrotropic cavity can induce quantum phase transition in the cavity embedded quantum ring leading to the step-like change in the total angular momentum of the system, which leads to the step-like dependence of the persistent current on the light-matter coupling. The effect is very similar to the Aharonov Bohm effect, but the magnetic flux is produced solely by the cavity vacuum fluctuations. We have shown how the transition can be probed via the weak probing illumination of the system and subsequent measurement of the statistical properties of the emitted light. The presented results suggest new approaches to cavity material engineering in the ultrastrong light-matter coupling regime with gyrotropic cavities.



\appendix

\section{Gyrotropic cavity}\label{app:gyrocac}
The optical cavity is formed by two lossless ferromagnetic mirrors with gyrotropic permittivity tensor~\cite{Tokatly_2021}
\begin{equation}
    \boldsymbol{\varepsilon} = \begin{bmatrix}
        \epsilon & i\epsilon_{H} & 0\\
        -i\epsilon_{H} & \epsilon & 0\\
        0 & 0 & \epsilon
    \end{bmatrix},\quad\epsilon<0,\quad\epsilon_{H}=\frac{4\pi\sigma_{H}}{\omega},
\end{equation}expressed in Gaussian units, which the convention adopted for this paper. In addition, $\hbar$ is shown explicitly for the sake of clarity in every appendix. A negative refraction occurs naturally in ferromagnets at microwave frequency~\cite{Engelbrecht_2011}, i.e. $\omega\sim1-100~\mathrm{GHz}$, which corresponds to the range of ferromagnetic resonance~\cite{Krupka_2018,Krupka_2020}. At optical frequency~\cite{Yang_2022} and beyond~\cite{Krupka_2021}, the non-diagonal elements of the permeability tensor $\boldsymbol{\mu}$, which has the form of a Polder tensor, become trivial, so that the material is effectively gyroelectric. In this case, the cavity modes satisfy
\begin{equation}\label{eq:cav_modes}
    \tan\left(\frac{\omega D}{2c}\right) = \sqrt{\lvert\epsilon\rvert\pm\frac{4\pi\sigma_{H}}{\omega}},
\end{equation}where $D$ is the length of the cavity (see Fig.~\ref{fig:cavity}). 

The previous equation is transcendental and possesses an infinite number of solutions, denoted $\omega_{\lambda}$ for a suitable label $\lambda$ and characterized by the handedness $\sigma=\pm1$ of the circular polarization (only when considering the two lowest modes can the modes be identified using $\sigma$). Since the degeneracy is lifted, the label $\lambda$ need not explicitly contain the polarization, hence
\begin{equation}
    \Op{H}_{\mathrm{cav}}=\sum_{\left(\lambda,\sigma\right)}\hbar\omega_{\lambda}\left(\Op{a}_{\lambda}^{\dagger}\Op{a}_{\lambda}+\frac{1}{2}\right),
\end{equation}where the creation ($\Op{a}^{\dagger}_{\lambda}$) and annihilation ($\Op{a}_{\lambda}$) operators satisfy the usual bosonic commutation relations. 

In the Coulomb gauge and long-wavelength limit~\cite{Rokaj_2022}, the vector potential operator in polar coordinates is $\Op{\mathbf{A}}=\Op{A}_{\rho}\mathbf{e}_{\rho}+\Op{A}_{\phi}\mathbf{e}_{\phi}$, with
\begin{equation}
    \Op{A}_{\phi} = \sum_{\left(\lambda,\sigma\right)}\frac{A_{\lambda}}{\sqrt{2}}i\sigma\left(\Op{a}_{\lambda}e^{i\sigma\phi} - \Op{a}_{\lambda}^{\dagger}e^{-i\sigma\phi}\right),\quad A_{\lambda} = \sqrt{\frac{2\pi\hbar c^{2}}{V\omega_{\lambda}}},
\end{equation}$\phi$ being the angular coordinate of the particle, and $\Op{A}_{\rho}=-\partial_{\phi}\Op{A}_{\phi}$.

\section{Effective Hamiltonian}\label{app:Heff}
The derivation of the effective Hamiltonian is obtained using all modes, without loss of generality. For an electron of effective mass $m^{\star}$ (simply $m$ in the main to avoid a cumbersome notation) and charge $-e<0$ embedded in a cavity subject to a parabolic circular potential centered at $\rho=R$
\begin{equation}
    \Op{H} = \frac{1}{2m^{\star}}\left(\Op{\mathbf{p}} + \frac{e}{c}\Op{\mathbf{A}}\right)^{2} + V\left(\Op{\mathbf{r}}\right) + \sum_{\lambda}\hbar\omega_{\lambda}\left(\Op{a}_{\lambda}^{\dagger}\Op{a}_{\lambda} + \tfrac{1}{2}\right),
\end{equation}where $\Op{\mathbf{p}}\to-i\hbar\boldsymbol{\nabla}$ and $V(\Op{\mathbf{r}})\to K(\rho-R)^{2}/2$, $K\in\mathbb{R}$, in coordinate space, and $\lambda$ is an appropriate label. By applying the transformations $\rho\to\rho/R$, $\Op{A}_{\rho,\phi}\to\Op{A}_{\rho,\phi}/A_{0}$ for some $A_{0}$ with vector potential dimensions, introducing the variables $g=eA_{0}R/(\hbar c)$, $k^{2}=KR^{4}m^{\star}/\hbar^{2}$, $\Tilde{\omega}_{\lambda}=\hbar\omega_{\lambda}/\frac{\hbar^{2}}{2m^{\star}R^{2}}$, and setting $E_{\mathrm{ring}}=\hbar^{2}/(2m^{\star}R^{2})\equiv1$ to make the Hamiltonian dimensionless normalized to $E_{\mathrm{ring}}$ reads
\begin{equation}
    \Op{H} = \left(\Op{\mathbf{p}} + g\Op{\mathbf{A}}\right)^{2} + V\left(\Op{\mathbf{r}}\right) + \sum_{\lambda}\Tilde{\omega}_{\lambda}\left(\Op{a}_{\lambda}^{\dagger}\Op{a}_{\lambda} + \frac{1}{2}\right),
\end{equation}where $\Op{\mathbf{p}}\to-i\boldsymbol{\nabla}$, $V(\Op{\mathbf{r}})\to k^{2}\left(\rho-1\right)^{2}$, and the last term is subsequently denoted as $\Op{H}_{\mathrm{ph}}$. Due to gauge ambiguities~\cite{DiStefano2019,Stokes_2024}, the truncation to a one-dimensional space must be carried out in the dipole gauge, obtained by applying a Power-Zienau-Woolley~\cite{Mandal_2023} (unitary) transformation $\Op{\Upsilon}$ such that
\begin{equation}\label{eq:PZWtrans}
    \Op{\Upsilon}^{\dagger} = \exp\left[-ig\mathbf{r}\cdot\Op{\mathbf{A}}\right] = \exp\left[-ig\rho\Op{A}_{\rho}\right] = \prod_{\left(\lambda,\sigma\right)}\Op{D}\left(\alpha_{\lambda,\sigma}\right),
\end{equation}where $\alpha_{\lambda,\sigma}=-ig\rho A_{\lambda}e^{-i\sigma\phi}/\sqrt{2}$ is the argument of the displacement operator. The result is derived using the Baker–Campbell–Hausdorff formula~\cite{Sakurai_2020}, the derivative of the exponential map~\cite{Rossmann_2002}, the commutation relations
\begin{equation}
    \comm{\Op{A}_{i}}{\Op{A}_{j}} = -i\epsilon_{ij}\sum_{\left(\lambda,\sigma\right)}\sigma A_{\lambda}^{2},
\end{equation}where $i,j=x,y$ or $\rho,\phi$, and $\epsilon_{ij}$ is the two-dimensional Levi-Civita symbol, and the usual relations between the displacement operators and the creation and annihilation operators. The final expression reads
\begin{equation}
    \begin{split}
        \Op{H} =& -\nabla^{2} -ig^{2}\partial_{\phi}\sum_{\left(\lambda,\sigma\right)}\sigma A_{\lambda}^{2} + \frac{1}{4}g^{4}\rho^{2}\left(\sum_{\left(\lambda,\sigma\right)}\sigma A_{\lambda}^{2}\right)^{2}\\
        &+ k^{2}\left(\rho-1\right)^{2}\\
        &+ \Op{H}_{\mathrm{ph}} + \sum_{\left(\lambda,\sigma\right)}\Tilde{\omega}_{\lambda}\left(\alpha_{\lambda,\sigma}\Op{a}_{\lambda}^{\dagger} + \alpha_{\lambda,\sigma}^{\ast}\Op{a}_{\lambda} + \frac{1}{2}g^{2}\rho^{2}A_{\lambda}^{2}\right).
    \end{split}
\end{equation}For $k\gg1$, the eigensystem for the radial part satisfies
\begin{equation}
    \left[-\partial_{\rho}^{2} - \frac{1}{\rho}\partial_{\rho} + k^{2}\left(\rho-1\right)^{2}\right]\chi\left(\rho\right) = \mathcal{E}_{\rho}\chi\left(\rho\right)
\end{equation}where $\mathcal{E}_{\rho}\in\mathbb{R}$. Setting $\chi(\rho)=\rho^{-1/2}P(\rho)$, defining a shifted coordinate $x=\rho-1$ and keeping only the highest order terms in $k$ and lowest order in $\lvert x\rvert\ll1$ remembering that $x\sim1/\sqrt{k}$ yields
\begin{equation}
    \left(-\partial_{x}^{2} + k^{2}x^{2}\right)P\left(x\right) = \mathcal{E}_{x}P\left(x\right),
\end{equation}where $\mathcal{E}_{x}=\mathcal{E}_{\rho}-1/4$, which corresponds to the quantum harmonic oscillator. The lowest energy solution is $P(x)=(k/\pi)^{1/4}e^{-kx^{2}/2}$, hence $\chi(\rho)=(k/\pi)^{1/4}e^{-k(\rho-1)^{2}/2}$, for $1/\sqrt{\rho}\approx1$ due to the strong confinement.

In this limit, $\rho\to1$ and the effective Hamiltonian after ignoring constant terms reads
\begin{equation}
    \begin{split}
        \Op{H}_{\mathrm{eff}} = -\partial_{\phi}^{2} -ig^{2}\partial_{\phi}\sum_{\left(\lambda,\sigma\right)}\sigma A_{\lambda}^{2} + \frac{1}{4}g^{4}\left(\sum_{\left(\lambda,\sigma\right)}\sigma A_{\lambda}^{2}\right)^{2}\\
        + \Op{H}_{\mathrm{ph}} + \sum_{\left(\lambda,\sigma\right)}\Tilde{\omega}_{\lambda}\left(ig\frac{A_{\lambda}}{\sqrt{2}}\left(\Op{a}_{\lambda}e^{i\sigma\phi} - \Op{a}_{\lambda}^{\dagger}e^{-i\sigma\phi}\right) + \frac{1}{2}g^{2}A_{\lambda}^{2}\right).
    \end{split}
\end{equation}Returning to the Coulomb gauge using the inverse of the transformation~\eqref{eq:PZWtrans} yields
\begin{equation}
    \Op{H} = \left(-i\partial_{\phi} + g\Op{A}_{\phi}\right)^{2} + \Op{H}_{\mathrm{ph}}.
\end{equation}Reintroducing the ring energy and the initial variables gives the expression in the main text.

\section{Single mode limit}\label{app:sml}
\begin{figure}
\includegraphics[width=\linewidth]{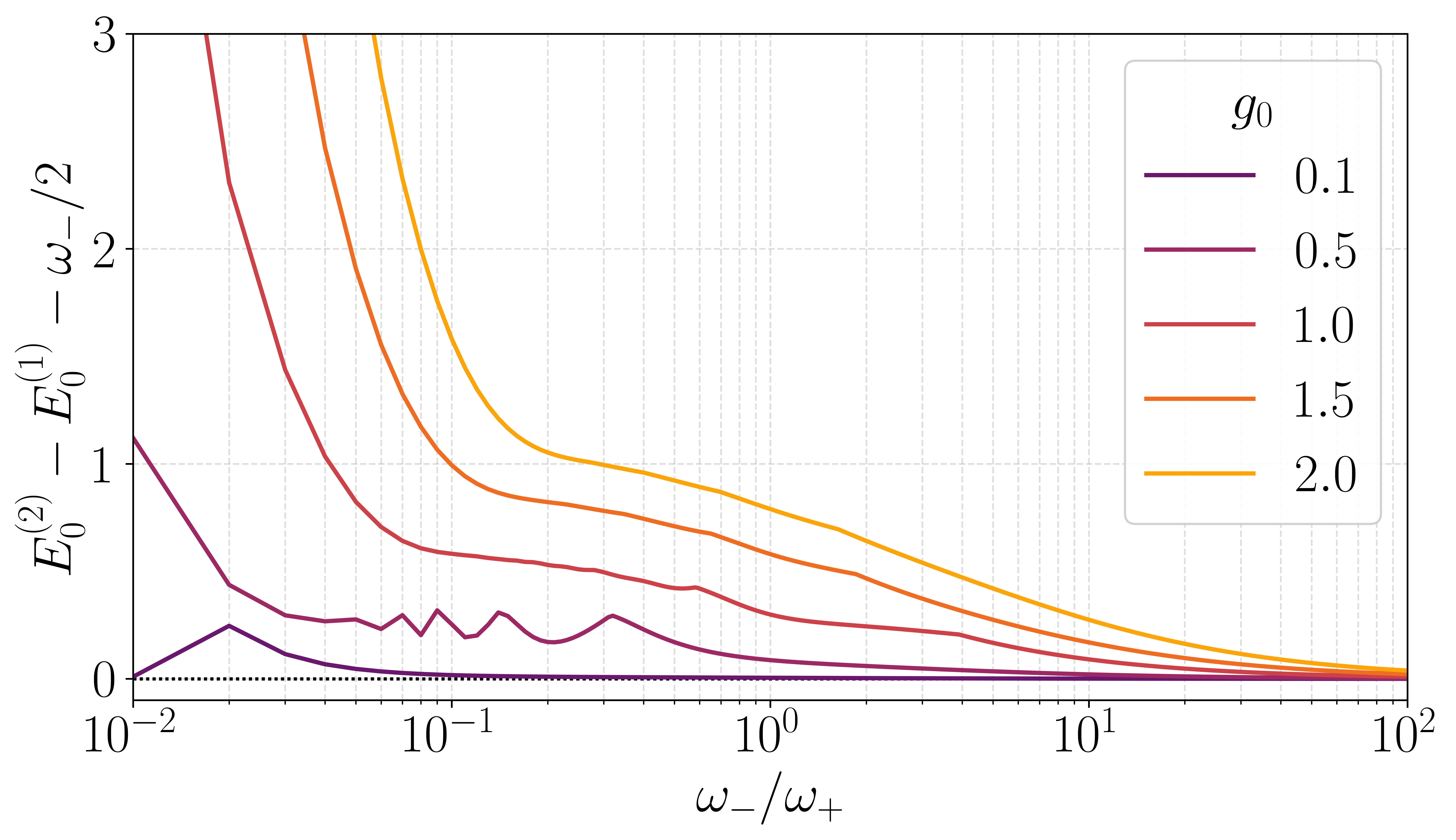}
\caption{\label{fig:SingleModeLimit}Single mode limit. $E_{0}^{(2)}$ ($E_{0}^{(1)}$) is the ground state energy for the double modes (single mode) Hamiltonian. The results were obtained for $\omega_{+}=7\pi/22\approx1$, $\sigma=+1$, and $-3\le j\le3$.}
\end{figure}

The Hamiltonian~\eqref{eq:Hprime} with $-i\partial_{\phi}\to j$ can be rewritten as
\begin{equation}
    \begin{split}
        \Op{H}^{\prime} &= \Op{H}^{\prime}_{+} + 2\left(g_{-}^{2}+\Op{n}_{-}\right)\left(j+g_{+}^{2}-\Op{n}_{+}\right) + \left(g_{-}^{2}+\Op{n}_{-}\right)^{2} + \\
        &\omega_{-}\left[(\Op{a}_{-}^{\dagger}-ig_{-})(\Op{a}_{-}+ig_{-})+\frac{1}{2}\right],
    \end{split}
\end{equation}where
\begin{equation}
    \Op{H}^{\prime}_{+} = \left(j+g_{+}^{2}-\Op{n}_{+}\right)^{2} + \omega_{+}\left[(\Op{a}_{+}^{\dagger}-ig_{+})(\Op{a}_{+}+ig_{+})+\frac{1}{2}\right],
\end{equation}is the single mode Hamiltonian for $\omega_{+}$. If one of the frequencies, say $\omega_{-}$, is much larger than the other, then $g_{-}\propto\omega_{-}^{-1/2}\to0$ and the ground state occurs for $n_{-}=0$, so 
\begin{equation}
    \Op{H}^{\prime}\simeq\Op{H}^{\prime}_{+} + \frac{\omega_{-}}{2},
\end{equation}where the last term only add a shift to the spectrum. The graphical result for the 2 modes versus 1 mode limit is shown in Fig.~\ref{fig:SingleModeLimit}.

%

\end{document}